\PassOptionsToPackage{svgnames,dvipsnames,table}{xcolor}
\documentclass[preprint, journal]{vgtc}            

\onlineid{4269}

\vgtccategory{Research}

\vgtcpapertype{please specify}

\title{SciDaSynth: Interactive Structured Data Extraction from \\Scientific Literature with Large Language Model}

\author{%
  Xingbo Wang\textsuperscript{†},
  Samantha L. Huey\textsuperscript{†},
  Rui Sheng,
  Saurabh Mehta\textsuperscript{*},
  Fei Wang\textsuperscript{*}
}

\authorfooter{
\item 
  \textsuperscript{†}: These authors contributed equally to this work;
  \textsuperscript{*}: Both are corresponding authors.
    \rev{\item Xingbo Wang is with Bosch Research North America \& Bosch Center for Artificial Intelligence (BCAI). This work is done when he is a postdoctoral associate at Weill Cornell Medicine. Email: wangxbzb@gmail.com.}
  \item Fei Wang is from Weill Cornell Medicine, Cornell University, New York, USA.
  Email: few2001@med.cornell.edu.
  \item Samantha L. Huey and Saurabh Mehta are from Cornell Joan Klein Jacobs Center for Precision Nutrition and Health, Cornell University, Ithaca, USA.
  E-mail: \{slhuey, smehta\}@cornell.edu

  \item Rui Sheng is with Hong Kong University of Science and Technology, Hong Kong.
  E-mail: rshengac@connect.ust.hk
}

\abstract{%
The explosion of scientific literature has made the efficient and accurate extraction of structured data a critical component for advancing scientific knowledge and supporting evidence-based decision-making.
However, existing tools often struggle to extract and structure multimodal, varied, and inconsistent information across documents into standardized formats.
We introduce SciDaSynth, a novel interactive system powered by large language models (LLMs) that automatically generates structured data tables according to users' queries by integrating information from diverse sources, including text, tables, and figures.
Furthermore, SciDaSynth supports efficient table data validation and refinement, featuring multi-faceted visual summaries and semantic grouping capabilities to resolve cross-document data inconsistencies. 
A within-subjects study with nutrition and NLP researchers demonstrates SciDaSynth's effectiveness in producing high-quality structured data more efficiently than baseline methods. We discuss design implications for human-AI collaborative systems supporting data extraction tasks.
}

\keywords{Data extraction, Large language models, Knowledge base, Scientific literature}

\graphicspath{{figs/}{figures/}{pictures/}{images/}{./}} 

\usepackage{tabu}                      
\usepackage{booktabs}                  
\usepackage{lipsum}                    
\usepackage{mwe}                       

\usepackage{mathptmx}                  
\usepackage{fontawesome}
\usepackage{comment}

\newcommand{\ie}{i.e.}
\newcommand{\eg}{e.g.}

\newcommand{\etal}{et al.}

\newcommand{\imp}[1]{\textbf{\textit{{#1}}}}

\newcommand{\name}{{\textit{SciDaSynth}}}
\newcommand{\rev}[1]{{\color{black} #1}}

\usepackage{inconsolata}
\usepackage{graphicx}
\usepackage{fontawesome}
\usepackage{comment}
\usepackage{fancyvrb}
\usepackage{multirow}

\RequirePackage[breakable,skins]{tcolorbox}

\begin{document}


\firstsection{Introduction}\label{sec.introduction}

\maketitle

The rapid advancement of scientific research has led to unprecedented growth in research literature across various disciplines.
Extracting and synthesizing structured knowledge from this vast information landscape has become increasingly crucial for advancing scientific understanding and informing evidence-based decision-making. 
Within this process, data extraction—the identification and structuring of relevant information from scientific literature—is a critical stage where efficiency and precision are paramount~\cite{Taylor88}, particularly in time-sensitive domains.
A relevant example is the early COVID-19 pandemic, where researchers urgently needed to determine the safety of breastfeeding for women with COVID-19~\cite{WHO2020BreastfeedingCOVID19}. This required the rapid and accurate extraction of data on experimental conditions (\eg, population demographics, study settings) and health outcomes from a rapidly expanding body of literature.

The structured data resulting from this process, often organized into tables, is essential for systematic comparison across studies, quantitative meta-analyses, and drawing comprehensive conclusions from diverse sources of evidence. Such data is crucial for organizations like the World Health Organization (WHO) in developing and disseminating timely, evidence-based guidelines~\cite{who2014}.

Despite its importance, data extraction remains a cognitively demanding and time-consuming task. 
Researchers often need to manually distill relevant information from multiple papers, switching between different documents and data entry tools. 
This process is not only inefficient but also prone to inconsistencies and errors, highlighting a critical need to streamline the data extraction process.
Addressing this need presents several challenges:
1) \textbf{Multimodal information in literature.}
Scientific papers often contain diverse modalities of information, such as text, tables, and figures.
The multimodality adds complexity to identifying the relevant information within each modality scattered throughout a paper and integrating it into a coherent and structured format.
2) \textbf{Variation and inconsistencies across literature.}
The style, structure, and presentation of the papers can significantly vary from one to another.
The variation and inconsistencies make it difficult to standardize the information.
For example, the same concepts may be described using different terminologies or measurement units.
3) \textbf{Flexibility and domain adaptation}.
Users may have varying research questions for a collection of papers, and these papers can span across different domains. 
Therefore, the system must be flexible enough to adapt to the diverse data needs of different users and domains.

Existing approaches to address these challenges have shown promise but face several key limitations. 
\rev{Tools and methods~\cite{papermage2023, scibert2019,evidencemap,nye2020trialstreamer,lehman2019inferring}}focusing on extracting keywords, tables, and figures from documents help narrow the scope of relevant information but lack the flexibility to accommodate diverse extraction needs. 
Users still need to manually explore, select, and integrate relevant data. 
\rev{Question-answering systems~\cite{chatgpt,claude}} have improved flexibility by allowing users to formulate information needs as queries about document content. 
However, these systems often produce unstructured text outputs, requiring significant effort to organize into desired structures. 
While \rev{some systems~\cite{elicit}} present information in tabular formats, they fall short in standardizing the data and resolving cross-document inconsistencies.

To address these limitations,
we present \name{}, an interactive system designed to empower researchers to efficiently and reliably extract and structure data from the scientific literature, \rev{especially after completing paper search and screening.}
\rev{By leveraging large language models (LLMs) within a retrieval-augmented generation framework (RAG)~\cite{lewis2020retrieval}}, the system interprets users' queries, extracts relevant information from diverse modalities in scientific documents, and generates structured tabular output.
\rev{Unlike standard prompting,  which relies solely on a model’s pretrained knowledge (and can be outdated due to the LLM’s training cutoff), RAG dynamically retrieves and integrates up‐to‐date, domain‐specific information into prompts. By injecting the retrieved information into the generation process, RAG reduces hallucinations and improves factual accuracy~\cite{ji2023survey}.}
To further ensure data accuracy, the system incorporates a user interface that establishes and maintains connections between the extracted data and the original literature sources, enabling users to iteratively validate, correct, and refine the data. Additionally, \name{} offers multi-faceted visual summaries of data dimensions and subsets, highlighting variations and inconsistencies across both qualitative and quantitative data. The system also supports flexible data grouping based on semantics and quantitative values, enabling users to standardize data by manipulating these groups and performing data coding or editing at the group level. In addition, follow-up query instructions can be applied to specific data groups for further refinement.
We conduct a within-subject study with researchers from nutrition and natural language processing (NLP) domains to evaluate the efficiency and accuracy of \name{} for data extraction from research literature.
The quantitative analyses show that using \name{}, participants could produce high-quality data in a much shorter time than the baseline methods.
Moreover, we discuss user perceived benefits and limitations.

In summary, our major contributions are:
\begin{itemize}
    \setlength{\itemsep}{0pt}
    \setlength{\parskip}{0pt}
    \item \name{}\footnote{\rev{System source code: \href{https://github.com/xingbow/SciDaEx}{https://github.com/xingbow/SciDaEx}}}, an interactive system that integrates LLMs to assist researchers in extracting and structuring multimodal scientific data from the extensive literature. 
    The system combines flexible data queries, multi-faceted visual summaries, and semantic grouping in a cohesive workflow, enabling efficient cross-document data validation, inconsistency resolution, and refinement. 
    \item The quantitative and qualitative results of our user study reveal the effectiveness and usability of \name{} for data extraction from the scientific literature.
    \item Implications for future system designs of human-AI interaction for data extraction and structuring.
\end{itemize}

\section{Related Work}\label{sec.related_work}

\subsection{Structured Information Extraction from Literature}
\label{rw.data_extraction}
The exponential growth of scientific papers has generated large-scale data resources for LLMs' building and applications for information extraction tasks, such as named entity recognition and relation extraction in scientific domains.
\rev{These models fall into two broad categories: encoder-only (non-generative) LLMs and generative models (autoregressive LLMs).
Encoder-only models (often called ``autoencoding'' models) are trained to produce compact vector representations of input text. These vectors are useful for downstream classification tasks~\cite{lewis2020}.
For example, SciBERT~\cite{scibert2019} uses the BERT architecture by pre-training on millions of scientific abstracts and full-text papers; it excels at classification, entity recognition, and retrieval tasks but is not good at generating new text. Variants like OpticalBERT and OpticalTable-SQA~\cite{zhao2023opticalbert} fine-tune the models on domain-specific corpora (\eg, biomedical or materials science) to boost performance on specialized extraction tasks.
Generative, or autoregressive, large language models (LLMs) predict the next word in a sequence, enabling them to create fluent text and even structured outputs directly from user prompts. Models such as GPT-4~\cite{achiam2023gpt} (and earlier InstructGPT) have hundreds of billions of parameters and have been further refined with techniques like instruction tuning and reinforcement learning from human feedback (RLHF). 
This training paradigm allows zero-shot or few-shot prompting: users can describe an extraction task in natural language and receive structured results—JSON, CSV—without any additional fine-tuning. 
In materials science, Dagdelen \etal~\cite{Dagdelen2024} demonstrated how GPT-4 could extract entities and relations and output them as JSON records with high fidelity.
In this paper, we chose GPT-4 series generative models over open-weight alternatives (\eg, Llama 3/4) for three key reasons. They are more reliable and accurate in following users' instructions (\eg, producing structured output) and handling complex, domain-specific queries.
Moreover, unlike open-weight models that usually need data-intensive fintuning for domain adaptation, commercial LLMs like GPT-4 work well for low-resource setting across diverse domains out-of-the-box with great flexibility.}

Besides, data in scientific literature is another particular focus for extraction.
The data is usually stored in tables and figures in PDFs of research papers,
and many toolkits are available to parse PDF documents, such as PaperMage~\cite{papermage2023}, GROBID~\cite{GROBID}, Adobe Extract API~\cite{adobeExtractText}, GERMINE~\cite{tkaczyk2015cermine}, GeoDeepShovel~\cite{geodeepshovel}, PDFFigures 2.0~\cite{clark2016pdffigures}. 
Here, we leverage the off-the-shelf tool to parse PDF text, tables, and figures.
Besides the tools in the research community, Elicit~\cite{elicit} is a commercial software that facilitates systematic review. It enables users to describe what data to be extracted and create a data column to organize the results.
However, it does not provide an overview of the extracted knowledge to help users handle variation and inconsistencies across different research literature.
Here, we also formulate the knowledge as structured data tables. Moreover, we provide multi-faceted visual and text summaries of the data tables to help users understand the research landscape, inspect nuances between different papers, and verify and refine the data tables interactively.

\subsection{Tools for Literature Reading and Comprehension}
\label{rw.literature_tools}
Research literature reading and comprehension is cognitively demanding, and many systems have been developed to facilitate this process~\cite{aug_sci_paper, PaperPlain, spotlight2016, scim2023, kang2022threddy, kang2023synergi, fok2023qlarify, tablelinking2018, marvista2023, PENG2022102898,jardim2022automating}.
One line of research studies aims to improve the comprehension and readability of individual research papers.
To reduce barriers to domain knowledge, ScholarPhi~\cite{aug_sci_paper} provided in-situ support for definitions of technical terms and symbols within scientific papers.
PaperPlain~\cite{PaperPlain} helped healthcare consumers to understand medical research papers by AI-generated questions and answers and in-situ text summaries of every section.
\rev{EvidenceMap~\cite{evidencemap} leverages three-level abstractions to support medical evidence comprehension.}
Some work~\cite{paperquest2016, chau2011apolo} designed interactive visualizations to summarize and group different papers and guide the exploration.
Some systems support fast skimming of paper content.
For example, Spotlight~\cite{spotlight2016} extracted visual salient objects in a paper and overlayed it on the top of the viewer when scrolling. Scim~\cite{scim2023} enabled faceted highlighting of salient paper content.
To support scholarly synthesis, Threddy~\cite{kang2022threddy} and Synergi~\cite{kang2023synergi} facilitated a personalized organization of research papers in threads. Synergi further synthesized research threads with hierarchical LLM-generated summaries to support sensemaking.
To address personalized information needs for a paper, Qlarify~\cite{fok2023qlarify} provided paper summaries by recursively expanding the abstract.
\rev{Some studies~\cite{jardim2022automating, marshall2016robotreviewer} developed and evaluated machine learning methods for risk of bias assessment in clinical trial reports.}
Although these systems help users digest research papers and distill knowledge with guidance, 
we take a step further by converting unstructured knowledge and research findings scattered within research papers into a structured data table with a standardized format.

\subsection{Document QA Systems for Information Seeking}
\label{rw.doc_qa}

People often express their information needs and interests in the documents using natural language questions~\cite{ter2020conversations}.
Many researchers have been working on building question-answering models and benchmarks~\cite{dasigi_dataset2021, Krithara2023, jin-etal-2019-pubmedqa, headqa2019, ruggeri-etal-2023-dataset} for scientific documents.
With recent breakthroughs in LLMs, some LLM-fused chatbots, such as ChatDoc~\cite{chatdoc}, ChatPDF~\cite{chatpdf}, ChatGPT~\cite{chatgpt}, Claude~\cite{claude}, are becoming increasingly popular for people to turn to when they have analytic needs for very long documents.
However, LLMs can produce unreliable answers, resulting in hallucinations~\cite{ji2023survey,khullar2024large}.
It is important to attribute the generated results with the source (or context) of the knowledge~\cite{commonsensevis}.
Then, automated algorithms or human raters can examine whether the reference source really supports the generated answers using different criteria
~\cite{gao-etal-2023-rarr, yue-etal-2023-automatic, measure_attirbuteLLM2023, bohnet2022attributed, menick2022teaching}.
In our work, we utilize retrieval-augmented generation techniques~\cite{lewis2020retrieval} to improve the reliability of LLM output by grounding it on the relevant supporting evidence in the source documents.
Then, we use quantitative metrics, such as context relevance, to evaluate the answer quality and prioritize users' attention on checking and fixing low-quality answers.

\section{Formative Study}\label{sec.design_process}
We aim to develop an interactive system that helps researchers distill, synthesize, and organize structured data from scientific literature in a systematic, efficient, and scalable way\footnote{Here, we focus on the stage where researchers have the final pool of studies ready for extraction, excluding literature search and screening.}.
To better understand the current practice and challenges they face during the process, we conducted a formative interview study.

\subsection{Participants and Procedures}\label{subsec.formative_users}
\subsubsection{Participants}
12 researchers (\imp{P1-P12}, five females, seven males, age: three from 18-24, nine from 25-34) were recruited from different disciplines, including medical and health sciences, computer science, social science, natural sciences, and mathematics.
Nine obtained PhD degrees and three were PhD researchers.
All of them had extracted data (\eg, interventions and outcomes) from literature, ten of which had further statistically analyzed data or narratively synthesized data.
Seven rated themselves as very experienced, where they had led or been involved with the extraction and synthesis of both quantitative and qualitative data across multiple types of reviews.
Five had expert levels of understanding and usage of computer technology for research purposes, and seven rated themselves at moderate levels.

\subsubsection{Procedures}
Before the interviews, we asked the participants to finish a pre-task survey, where we collected their demographics, experience with literature data extraction and synthesis, and understanding and usage of computer technology.
Then, we conducted 50-minute interviews with individuals over Zoom.
During the interviews, we inquired the participants about 
(1) their general workflow for data extraction from literature, 
desired organization format of data;
(2) what tools were used for data extraction and synthesis, and what are their limitations;
(3) expectations and concerns about computer and AI support.

\subsection{Findings and Discussions}\label{subsec.findings}

\subsubsection{Workflow and tools.}
After getting the final pool of included papers, participants first created a data extraction form (\eg, fields) to capture relevant information related to their research questions, such as data, methods, interventions, and outcomes. 
Then, they went through individual papers, starting with a high-level review of the title and abstract. Afterward, participants manually distilled and synthesized the relevant information required on the form.
The data synthesis process often involved iterative refinement, where participants might go back and forth between different papers to update the extraction form or refine previous extraction results.

Common tools used by participants included Excel (9/12) and Covidence or Revman (4/12) for organizing forms and results of data extraction.  Some participants also used additional tools like Typora, Notion, Python or MATLAB for more specialized tasks or to enhance data organization.
The final output of this process was structured data tables in CSV and XLSX format that provided a comprehensive representation of the knowledge extracted from the literature.


\subsubsection{Challenges}

\textit{Time-consuming to manually retrieve and summarize relevant data within the literature.}
Participants found it time-consuming to extract different types of data, including both qualitative and quantitative data, located at different parts of the papers, such as text snippets, figures, and tables.
\imp{P1} commented, ``Sometimes, numbers and their units are separated out at different places.''
The time cost further increases when facing ``many papers'' (7/12) to be viewed, ``long papers'' (5/12), or papers targeting very specialized domains they are not so familiar with (5/12).  
\imp{P3} added, ``When information is not explicit, such as limitations, I need to do reasoning myself.''
\imp{P5} said, ``It takes much time for me to understand, summarize, and categorize qualitative results and findings.''

\textit{Tedious and repetitive manual data entry from literature to data tables.}
After locating the facts and relevant information, participants need to manually input them into the data tables, which is quite low-efficiency and tedious.
\imp{P3} pointed out, ``... the data is in a table (of a paper), I need to memorize the numbers, then switch to Excel and manually log it, which is not efficient and can cause errors.''
\imp{P4} echoed, ``Switching between literature and tools to log data is tedious, especially when dealing with a large number of papers, which is exhausting.''

\textit{Significant workload to resolve data inconsistencies and variations across the literature.}
Almost all participants mentioned the great challenges of handling inconsistencies and variations in data, such as terminologies, abbreviations, measurement units, and experiment conditions, across multiple papers.
It was hard for them to standardize the language expressions and quantitative measurements.
\imp{P7} stated, ``Papers may not use the same terms, but they essentially describe the same things. And it takes me lots of time to figure out the groupings of papers.''
\imp{P9} said, ``I always struggle with choosing what words to categorize papers or how to consolidate the extracted information.''

\textit{Inconvenient to maintain connections between extracted data and the origins in literature.}
The process of data extraction and synthesis often required iterative review and refinement, such as resolving uncertainties and addressing missing information by revisiting original sources. 
However, when dealing with numerous papers and various types of information, the links between the data and their sources can easily be lost. 
Participants commonly relied on memory to navigate specific parts of papers containing the data, which is inefficient, unscalable, and error-prone. 
\imp{P8} admitted, ``I can easily forget where I extract the data from. Then, I need to do all over again.''

\subsubsection{Expectations and concerns about AI support}
Participants anticipated that AI systems could automatically extract relevant data from literature based on their requests (7/12), and organize it into tables (9/12).
They desired quick data summaries and standardization to (6/12) facilitate synthesis. 
Additionally, they wanted support for the categorization of papers based on user-defined criteria (4/12) and enabling efficient review and editing in batches (4/12). 
Besides, participants expected that the computer support should be easy to learn and flexibly adapt to their data needs.
Many participants stated that the existing tools like Covidence and Revman were somewhat complex, especially for new users who may struggle to understand their functionalities and interface interactions.

Due to the intricate nature of scientific research studies, participants shared concerns about the accuracy and reliability of AI-generated results.
They worried that AI lacks sufficient domain knowledge, and may generate results based on the wrong tables/text/figures.
\imp{P12} demanded that AI systems should highlight uncertain and missing information.
Many participants requested validation of AI results.


\subsection{Design Goals}\label{subsec.design_requirements}

Based on the current practice and challenges identified in our formative study and the specific needs of researchers engaged in data extraction, we distilled the following design goals:

\begin{itemize}
    \item \textbf{DG1. Support flexible and comprehensive data extraction and structuring.}  
    The system should enable users to customize data extraction queries for diverse data dimensions and measures. To reduce manual effort, it should automate the extraction of both qualitative and quantitative data from various modalities such as text, tables, and figures. The extracted data should be organized into structured tables, providing a solid foundation for further refinement and analysis.

    \item \textbf{DG2. Enable multi-faceted data summarization and standardization.}  
    To address inconsistencies and variations across literature, the system should provide an overview of key patterns and discrepancies in the extracted data regarding different dimensions and measures. It should also assist in standardizing data derived across multiple documents, such as terminologies, measurements, and categorizations.

    \item \textbf{DG3. Support efficient data validation and refinement.}  
    The system should address the need for ensuring the accuracy and reliability of extracted data:
    \begin{itemize}
        \item[\textbf{3.1}] \textbf{Provide preliminary evaluation of automated extracted data.} This helps users identify data errors and prioritize effort in validation.
        \item[\textbf{3.2}] \textbf{Facilitate easy comparison of extracted data against original sources.} This enables users to trace data origins to verify data accuracy.
        \item[\textbf{3.3}] \textbf{Enable efficient batch editing and refinement of data.} It should support data entry for data subsets (\eg, with the same dimension values).
    \end{itemize}
\end{itemize}

\section{System}\label{sec.system}
\begin{figure*}[!htb]
    \centering
    \includegraphics[width=\textwidth]{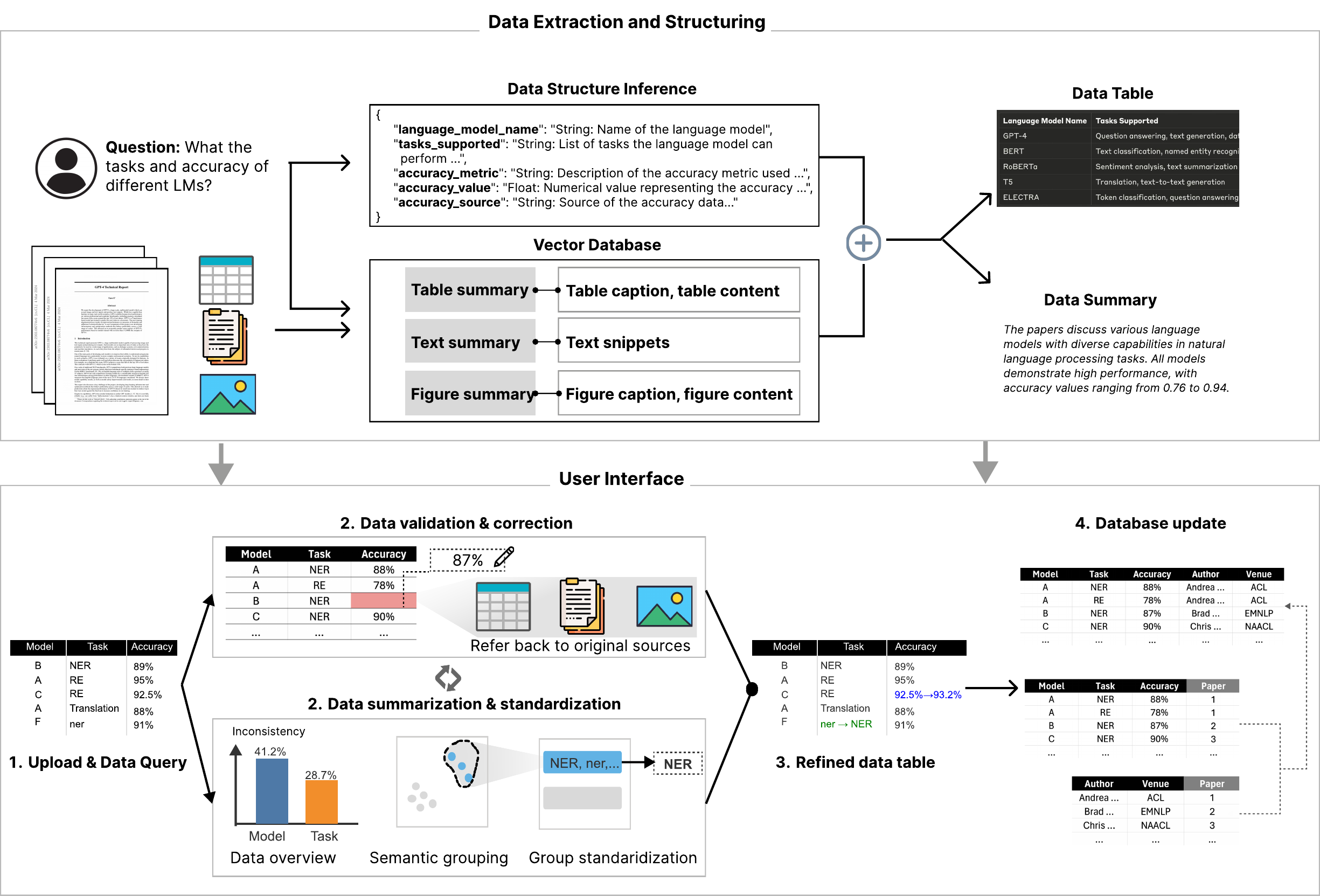}
    \caption{System workflow of \name{}: (1) Retrieval augmented generation (RAG) based technical framework for extracting and structuring data from figures, text, and tables in scientific documents using LLMs.
    (2) The user interface then allows for data extraction via question-answering, data validation, correction, summarization, standardization, and database updates through an iterative refinement process.
    }
    \label{fig:qa_framework}
\end{figure*}

Here, we introduce the design and implementation of \name{}.
First, we provide an overview of the system workflow (in \autoref{fig:qa_framework}). 
Then, we describe the technical pipeline of data extraction and structuring.
Finally, we elaborate the user interface designs and interactions.

\subsection{System Workflow}\label{sec.sys_overview}

After uploading the PDF files of research literature, users can interact with \name{} with natural language questions (\eg, \textit{``What are the task and accuracy of different LMs?''}) or customized data extraction forms in the chat interface (\autoref{fig:qa_framework}).
The system then processes this question and presents the user with a text summary and a structured data table (\imp{DG2}).
This interaction directly addresses the data needs without requiring tedious interface drag and drop (\imp{DG1}).

\begin{figure*}[!bht]
  \includegraphics[width=\textwidth]{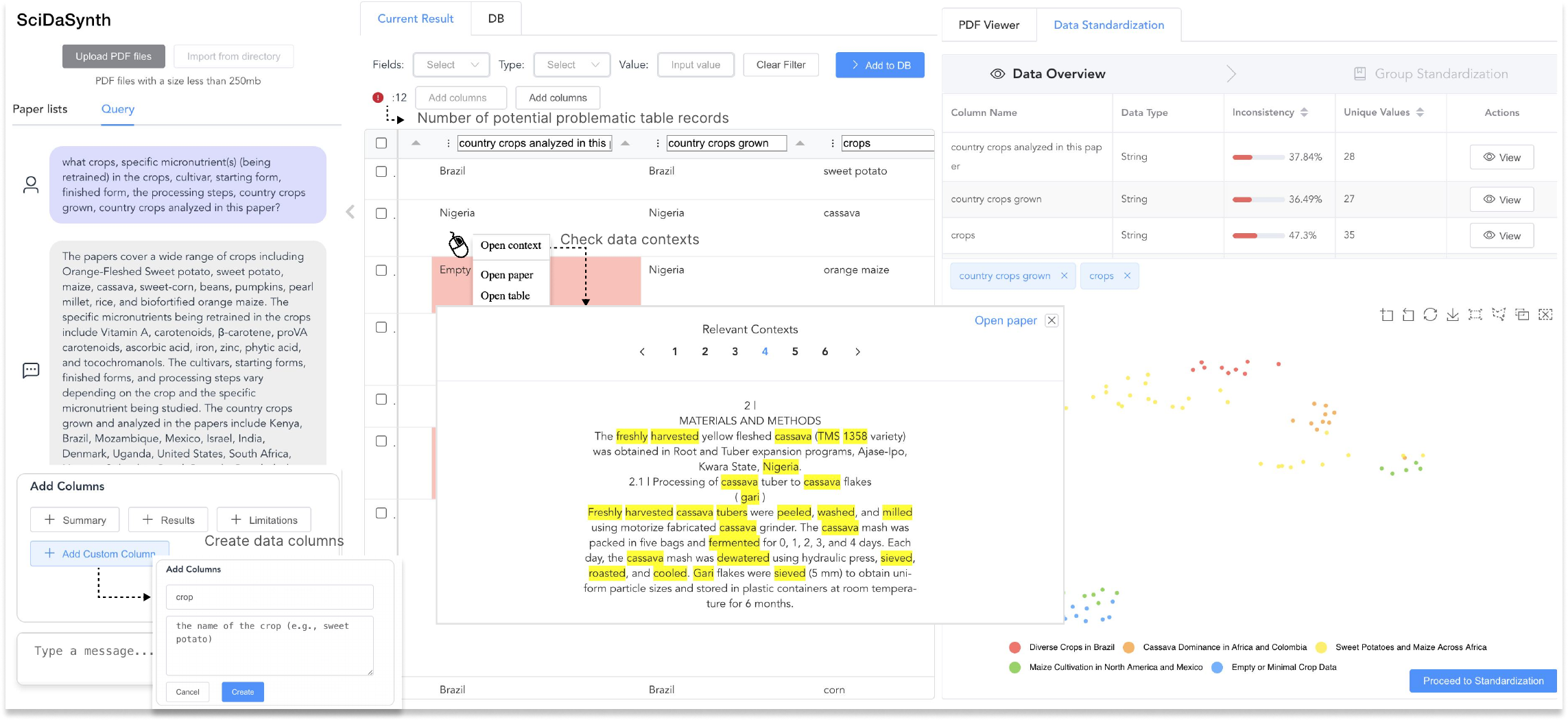}
  \caption{User interface of \name{}.
  The interfaces features: (A) A query panel for users to input natural language questions or select specific data attributes; 
  (B) A data table displaying extracted information with highlighting of potentially problematic records; 
  (C) Context menu options to validate data by examining relevant document snippets; 
  (D) PDF viewer for accessing original sources; 
  (E) Data standardization panel with multi-level and multi-faceted data summarization and standardization support.
  }
  \label{fig:teaser}
\end{figure*}

The data table provided by \name{} includes specific dimensions related to the user's question, such as \textit{``Model''}, \textit{``Task''}, and \textit{``Accuracy''}, along with corresponding values extracted from the literature. 
To guide users' attention to areas needing validation, the system highlights missing values ("Empty" cells) and records with low relevance scores (\imp{DGs 3.1, 3.2}).
To validate and refine data records, users can view the relevant contexts used by the LLM, with important text spans highlighted (\autoref{fig:teaser}-middle). They can also access the original PDF. 

To handle data inconsistencies across papers, the system provides multi-level and multi-faceted data standardization interface (\imp{DG2}).
First, users can gain an overview of data attributes and their consistency information (\autoref{fig:teaser}-right). 
Upon selecting specific attributes, the system performs semantic grouping of attribute values to help users identify contextual patterns and distributions of potential inconsistencies (\eg, full form vs. abbreviation). 
Next, based on the grouped attribute values and their visual summary, users can create, modify, rename, or merge the groups, effectively categorizing the data. Once satisfied with their groupings, users can apply standardization results to instantly update the main data table (\imp{DG3.3}). Throughout this process, the system provides real-time updates to charts and statistics to show the impact of standardization efforts. 

Once satisfied with the data quality, users can add the table to the database, where it's automatically merged with existing data. This process can be repeated with new queries to incrementally build a comprehensive database (\autoref{fig:teaser}-middle). Finally, users can export the entire database in CSV format for further analysis or reporting.

\subsection{Data Extraction and Structuring}\label{sec.data_extract}
We leverage LLMs~\footnote{We use GPT-4o for figure insights generation and table parsing, table structure inference, and data generation, GPT3.5 for text summarization, and text-embedding-small for vectorization. The choices are made considering speed, cost, and task complexity.} to extract and structure data from scientific literature based on user questions (\imp{DG1}). To mitigate hallucination issues and facilitate user validation of LLM-generated answers, we adopt the retrieval-augmented generation (RAG) framework by grounding LLMs in relevant information from the papers (as shown in ~\autoref{fig:qa_framework}).

The process begins with parsing the paper PDF collection into tables, text snippets (e.g., segmented by length and sections), and figures using a state-of-the-art toolkit for processing scientific papers~\cite{papermage2023}. To enhance the process, we generate data insights embedded in the figures using large vision-language models \rev{(\ie, GPT-4o in this paper)}. Due to the complexity of modern table structures, we also integrate LLMs to infer, parse out, and standardize the table structure (to CSV format) from table strings in PDF texts via few-shot prompting (details in Supplementary A).

\textbf{Vector database construction.}
We construct a vector database for efficient retrieval and question-answering using the processed text, figures, and tables. For each component, we generate a concise text summary using LLMs to index the original verbose content. This approach reduces noise and improves RAG quality. The text summary is then transformed into embeddings.

\textbf{Question-based retrieval.}
When a user poses a question, we encode it as a vector and perform a similarity-based search to identify relevant original content by comparing it with the summary vectors of figures, tables, and text snippets in the database.
The retrieved original content is later used for LLM's generation.

\textbf{Data table structure inference.}
Based on the user's question, we prompt LLMs to infer and design the structure of the data tables, including column names and value descriptions. This guides the LLM to formulate scoped, consistent, and standardized responses across different papers.

\textbf{Data extraction and structuring}
The user question, retrieved document snippets, and the inferred data structure are fed into LLMs to produce the final data table and an associated summary. We instruct the LLMs to answer questions solely based on the provided contexts and to output "Empty" for values that cannot be determined from the given information. The organization of structured data tables facilitates convenient human validation and refinement (\imp{DG3}).

\textbf{Data quality evaluation}
To assess the quality of the RAG process, we implement several unsupervised metrics computed by LLMs to judge answer quality regarding retrieved contexts and original questions~\cite{es-etal-2024-ragas,yu2024evaluationretrievalaugmentedgenerationsurvey}: answer relevancy measures how well the answer addresses the user's question; context relevancy evaluates the pertinence of the retrieved context to the question; and faithfulness assesses the degree to which the answer can be justified by the retrieved context. Additionally, we compute data missingness by tracking the proportion of ``Empty'' values in the generated table. This metric alerts users to insufficient or missing information in the original source (\imp{DG3.1}).

\begin{tcolorbox}[
  title=Data Table Structure Inference Example,
  fonttitle=\bfseries\small,
  colback=white,
  colframe=black,
  boxrule=0.5pt,
  left=2pt,
  right=2pt,
  top=2pt,
  bottom=2pt,
  arc=0pt,
  outer arc=0pt,
  fontupper=\small
]
\textbf{Example:}

\texttt{Q: What are the tasks and accuracy of different LMs?}
\vspace{0.5em}
\textbf{Inferred structure:}
\begin{Verbatim}[commandchars=\\\{\}]
\{\textbf{"language_model_name"}: "String: Name of the language model (e.g., 
 GPT-3, BERT)",
 \textbf{"tasks_supported"}: "String: List of tasks the language model can 
 perform (e.g., text generation, summarization, translation)",
 \textbf{"accuracy_metric"}: "String: Description of the accuracy metric used 
 (e.g., F1, BLEU)",
 \textbf{"accuracy_value"}: "Float: Numerical value for the accuracy (0-100)",
 \textbf{"accuracy_source"}: "String: Source of the accuracy data (e.g., 
 research paper, benchmark test)"\}
\end{Verbatim}
\end{tcolorbox}

\subsection{User Interface}\label{subsec.ui}

Building upon the RAG-based technical framework, the user interface streamlines the data extraction, structuring, refinement from scientific documents.
Based on the LLM-generated data tables, 
users can perform iterative data validation and refinement by pinpointing and correcting error-prone data records and by resolving data inconsistencies via flexible data grouping regarding specific attributes.
Finally, users can add the quality-ready data tables into the database.
Here, we will introduce the system designs and interactions following user workflow.

\subsubsection{Flexibly specify information needs}
Upon uploading the scientific documents into \name{}, users can formulate their questions in the Query tab by typing the natural language questions in the chat input (\imp{DG1}, \autoref{fig:teaser}).
Alternatively, users can select and add specific data attributes and their detailed explanations and requirements in a form box to query some complex terminologies. This form includes suggested starting queries such as study summary, results, and limitations.
Then, the system will respond to users' questions with a text summary and present a structured data table in the ``Current Result'' tab (\imp{DG2}).

\subsubsection{Guided data validation and refinement}
To make users aware of and validate the potentially problematic results (\imp{DG3.1}), \name{} highlights error-prone data records and values.
Specifically, the system highlights ``Empty'' cells and the error-prone table records (\textcolor{black}{\faInfoCircle{}} with a counter at top) using unsupervised metrics (in \autoref{sec.data_extract}).
Then, users can right click the specific table records or cells to access a context menu, which provides the option to view the relevant contexts used by the LLMs for data generation. Within these contexts, important text spans \rev{that exactly match} the generated data are highlighted for quick reference. If users find the contexts irrelevant or suspect LLM hallucination, they can easily access the original PDF content or parsed figures and tables in the right panel (\autoref{fig:teaser}) for verification (\imp{DG3.2}).
After identifying the errors, users can double click cells to edit the value and clear alerts by clicking \textcolor{black}{\faInfoCircle{}} in the rows.

\subsubsection{Multi-level and multi-faceted data summarization and standardization}
To resolve data inconsistencies across different literature, the system first present an overview of data attributes, data types, and inconsistency\footnote{measured by proportion of unique values in a attribute.} information (\autoref{fig:teaser}-right).
\par{}\textbf{Dimension-guided data exploration.} (\imp{DG2})
After selecting data attributes, the system performs visual semantic data grouping based on attribute values.
Specifically, each record (row) of the selected attributes in the table is transformed into a text description (``column\_name: value''), encoded as a vector, and projected on the 2D plane as a dot\footnote{Numerical values are converted into categories from ``low'' to ``high''.} where size correlates with frequency.
Users can hover individual dots to see the column values and their group labels. In addition, they can select a group of dots to examine full data records in the ``Current Result'' table.
The variations and similarities of dimension values for different rows are reflected in the distribution of clusters of dots using KMeans clustering.
To concretize the data variations, each cluster is associated with a text label generated by LLMs' summarization.
For example, the scatter plot groups ``crops'' values into colored clusters, such as sweet potatoes and maize.
Besides, users can select multiple attributes at once, such as ``nutrient value'' and ``measurement units'', to gain contextual insights into discrepancies into measurement units.

\par
\begin{figure}[!htb]
  \includegraphics[width=\columnwidth]{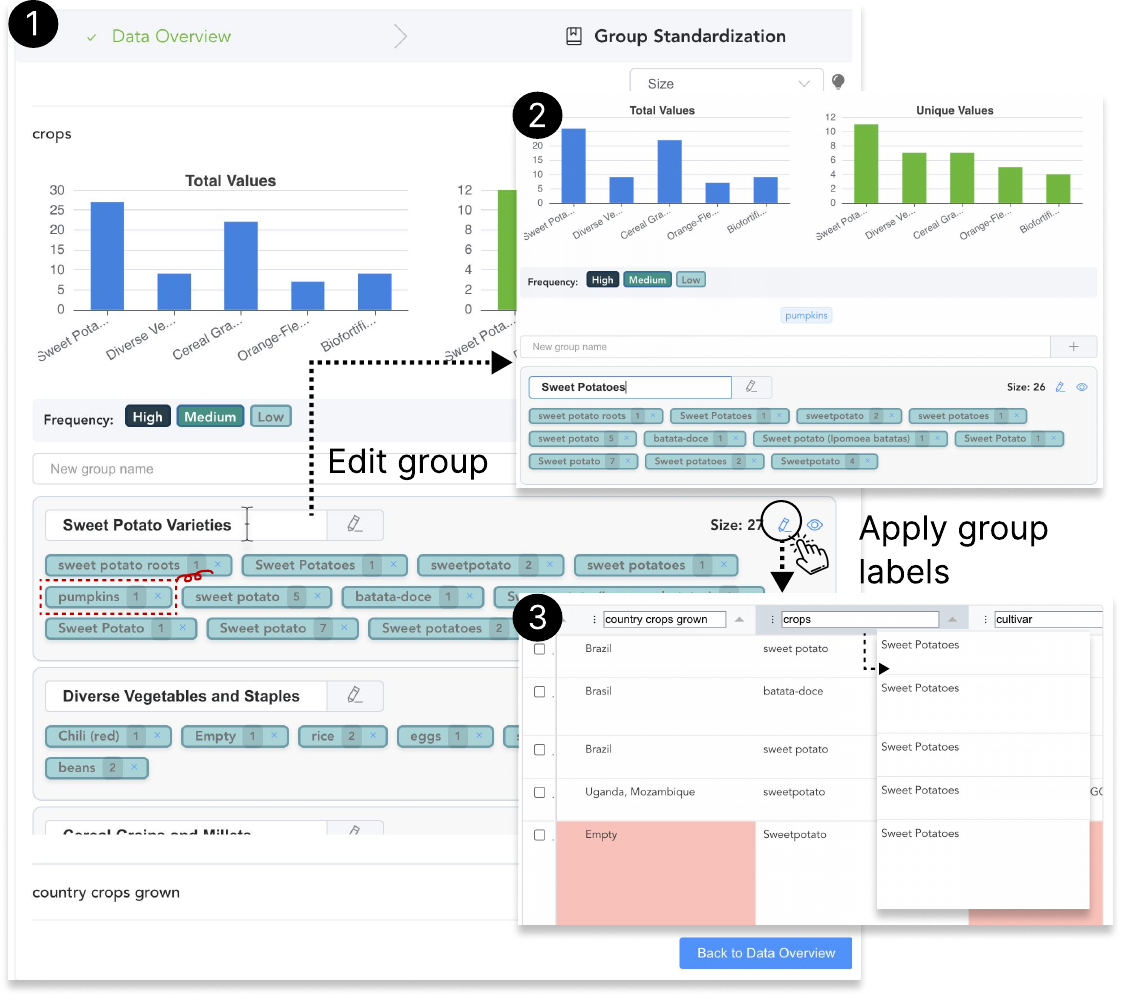}
  \vspace{-5mm}
  \caption{
   Group standardization process. Users start with major groups' statistics within a selected data attributes. Then, they can edit individual groups by changing the group labels and removing irrelevant values. Finally, they can apply edited group results to the data table.}
  \label{fig:data_standardization}
\end{figure}

{}\textbf{Group-based standardization.} (\imp{DG2})
After developing some high-level understanding of data variations, users can proceed to standardization of the selected data attributes (in \autoref{fig:data_standardization}).
Users can start with an overview of total and unique values for each major group within the attribute, visualized through bar charts. 
Below the charts, the system displays individual group cards, each representing a cluster of similar values. These cards are color-coded based on the frequency of occurrences (high, medium, low), allowing users to quickly identify prevalent and rare entries. Within each card, the system lists all unique value variations, along with their frequency counts. This granular view enables users to easily spot inconsistencies, misspellings, or variations in terminology.

The interface supports the following interactive standardization:
\begin{itemize}
    \setlength{\itemsep}{0pt}
    \setlength{\parskip}{0pt}
    \item Users can create new groups or rename existing ones to better categorize the data.
    \item Users drag-and-drop individual value entries between groups, facilitating the consolidation of similar terms.
    \item Inline editing tools enable users to modify group names or individual value entries directly.
    \item For individual group card, users can apply the standardization to the data table by clicking \textsf{\small\faicon{pencil}}. The results can be tracked and viewed by clicking \textsf{\small\faicon{eye}}.
\end{itemize}
As users make changes, the system provides real-time updates to the overview charts and group statistics, offering immediate feedback on the impact of standardization efforts.

\subsubsection{Iterative table editing and database construction}
When satisfied with the quality of the current table, users can add it to the database, where the system automatically merges it with existing data using outer joins on document names. This process can be repeated with new queries, allowing for the incremental construction of a comprehensive database. Once the data extraction is complete, users can download the entire database in CSV format for further analysis.

\section{Evaluation Design}\label{sec.evaluation}
We employed a within-subjects design where participants used both \name{} and a baseline system to extract data from two paper collections. The study aimed to answer the following research questions:

\begin{itemize}
    \setlength{\itemsep}{0pt}
    \setlength{\parskip}{0pt}
    \item \textbf{Effectiveness of data extraction}:
    \begin{itemize}
        \setlength{\itemsep}{0pt}
        \setlength{\parskip}{0pt}
        \item \textbf{Data quality:} How does \name{} impact the quality of extracted data?
        \item \textbf{Efficiency:} How does \name{} affect the speed of data extraction?
    \end{itemize}
    \item \textbf{User perceptions}: What are the perceived benefits and limitations of system designs and workflows?
\end{itemize}

\subsection{Experiment Settings}

\subsubsection{Participants}
We recruited a diverse group of 24 researchers:

\begin{itemize}
    \setlength{\itemsep}{0pt}
    \setlength{\parskip}{0pt}
\item \textbf{Group A: Nutrition Science} (12 participants; P1-P12)
This group (8 females, 4 males, aged 18-44) specialized in nutritional sciences, including food science and technology, human nutrition, medical and health sciences, and life sciences. They were five postdoctoral fellows and seven PhD students, all actively engaged in research.  Their technical expertise varied: five were expert users who regularly coded and programmed, while seven were intermediate users who coded as needed.

\item \textbf{Group B: NLP/ML Researchers} (8 participants; P13-P20)
This group (5 males, 3 females, aged 18-33) included researchers in natural language processing, machine learning, and artificial intelligence. There were 2 postdocs, 5 PhD students, and 1 MPhil student. They were expert in computer programming.

\end{itemize}

All participants were familiar with the target dataset dimensions through research or literature dimensions. 
They had experience in data extraction and synthesis for research studies. 

\subsection{Datasets}
We collected and processed two domain datasets based on the recent survey papers. The surveys cover diverse paper types and formats, such as randomized trial, peer-reviewed articles, meeting abstract and posters, and doctoral theses:
\begin{itemize}
    \item Dataset I (Nutrition Science) is based on a systematic review in Nature Food~\cite{Huey2023}, focusing on micronutrient retention in biofortified crops through various processing methods.
    \item  Dataset II (Large Language Models) is derived from a recent LLM survey~\cite{zhao2023surveylargelanguagemodels}, covering various aspects of LLM development and applications.
\end{itemize}
\rev{To build each dataset, we conveniently sampled 20 publications out of the original pool of included papers in each corresponding review. These papers had sufficient length and complexity and incorporated different modalities of information (text, tables, and figures) (see \autoref{tab:datasets_stats}), while also maintaining a manageable level of task complexity for the user study.}
After that, we pre-processed them as described in \autoref{sec.data_extract}.
The corresponding data tables from the original reviews served as ground truth for our evaluation.

\begin{table}[h]
\centering
\caption{Statistics of papers in Datasets \texttt{I} and \texttt{II}. Values are presented as mean (standard deviation). }
\label{tab:datasets_stats}
\resizebox{\columnwidth}{!}{%
\begin{tabular}{|l|c|c|c|c|}
\hline
& {Page \#} & {Character \#} & {Figure \#} & {Table \#} \\
\hline
Dataset I & 9.40 (2.97) & 34842 (14569) & 3.6 (1.95) & 5.2 (3.35) \\
\hline
Dataset II & 24.10 (7.50) & 74882 (22077) & 7.00 (3.97) & 13.60 (4.30) \\
\hline
\end{tabular}
}
\end{table}

\subsection{Baseline Implementation}
\textbf{Baseline A (Human)}
A simplified version of \name{} without automated data extraction or structuring, designed to replicate current manual practices. It includes (1) A PDF viewer with annotation, highlighting, and searching capabilities;
(2) Automatic parsing of paper metadata, tables, and figures;
(3) A data entry interface for manual table creation;
(4) Side-by-side views of PDFs and data tables;
This baseline allows us to assess the impact of \name{}'s automated features and standardization support on efficiency and data quality.

\textbf{Baseline B (Auomated GPT).}
We developed a fully automated system based on GPT-3.5/4, to generate data tables according to specified data dimensions. 
This baseline was intended to evaluate the accuracy of our technical framework for automatic data table generation. 
The implementation followed the data extraction and structuring approach of \name{} (described in \autoref{sec.data_extract}). 
To produce the data tables for comparison, we input the data attributes and their descriptions in JSON format as queries into the system and generate two data tables for the two splits (\ie, four data points) of each dataset.

\subsection{Tasks}
We designed tasks to simulate real-world data extraction scenarios while allowing for controlled evaluation across two distinct domains. Each participant was assigned to work with one complete dataset (either Dataset I or Dataset II), consisting of 20 papers in total.

\par{}Nutrition science researchers (P1-P12) worked with Dataset I, extracting ``crops (types)'', ``micronutrients (being retained)'', ``absolute nutrient raw value'',  and ``raw value measurement units''.

\par{} NLP/ML researchers (P13-P20) worked with Dataset II, extracting ``model name'', ``model size'', ``pretrained data scale'', ``hardware specifications (GPU/TPU)''.

These dimensions covered both qualitative and quantitative measurements, requiring engagement with various parts of the papers.

\rev{To ensure a robust comparison between the two interactive systems (\name{} and Baseline A (manual system)), we split each dataset into two subsets of 10 papers each, ensuring balanced distribution of paper characteristics such as paper length, number of figures and tables. 
Each participant extracted all four data dimensions for every paper in their assigned dataset (I or II) using both systems. 
This means that each participant used both systems for both subsets, producing two data tables in total.
Baseline B (automated baseline) was not operated by participants; it was separately run to compare automated performance against human-in-the-loop systems.}
\rev{To mitigate the ordering effects (\eg, learning effect), we counterbalanced the order of system usage (\name{} first or Baseline A first) and dataset splits (Split 1 first or Split 2 first), resulting in a 2 (system order) x 2 (data split order) within-subject design.}
Participants organized the extracted data into tables and downloaded them from the systems. The scenario was framed as ``working with colleagues to conduct a systematic review.''

\par{}Following the structured tasks, participants engaged in an \textbf{open-ended exploration} of the whole dataset with \name{}, allowing for insights into the system's full capacity in research use.

\subsection{Procedure}
We conducted the experiment remotely via Zoom, with both Baseline A and \name{} deployed on a cloud server. The study followed a structured procedure: pre-study setup and briefing (10 minutes), system tutorials (10 minutes each), main tasks with both systems and datasets (counterbalanced ordered), post-task surveys, free exploration of \name{} (15 minutes), and a concluding interview.

Participants first provided consent and background information. They then received tutorials on each system before performing data extraction tasks. After completing structured tasks with both systems, participants freely explored \name{} while thinking aloud. The study concluded with a semi-structured interview gathering feedback on system designs, workflow, and potential use cases.

Each session lasted approximately 2.5 hours, with participants compensated \$30 USD. This procedure enabled collection of comprehensive quantitative and qualitative data on \name{}'s performance and user experience across different research domains

\subsection{Measurements}\label{subsec.measure}
We evaluated \name{} using quantitative and qualitative measures focusing on effectiveness, efficiency, and user perceptions, with separate analyses for Datasets I and II.

\textbf{Effectiveness of data extraction} was assessed by evaluating the data quality and task completion time.
For \textit{data quality}, we compared the data tables generated by participants using \name{}, Baseline \texttt{A}, and the automated GPT baseline (Baseline \texttt{B}) against the original data tables from the systematic review. 
\rev{For each dataset, two expert raters who were blind to the system conditions independently scored the 3-point scale}: 0 (Not Correct), 1 (Partially Correct)\footnote{Records were generally correct but incomplete, missing some information for certain dimensions}, and 2 (Correct), based on accuracy and completeness.
\rev{The inter-rater agreement on individual dimensions is measured by Cohen's $\kappa$.
Generally, two raters had good agreement (>0.7) (as shown in \autoref{tab:kappa_summary}).
Disagreements were resolved through discussion to reach consensus scores.}
For \name{} and Baseline \texttt{A}, we calculated participants' scores for the corresponding dataset (one per participant, each ranging from 0 to 20).
Then, the paired Student's t-test was performed to compare the average scores of \name{} and Baseline \texttt{A}.
Baseline \texttt{B} yielded 2 scores per dataset split, compared using Mann-Whitney U tests~\cite{kang2023synergi}.

For \textit{task efficiency},  we measured task completion time from the moment the PDFs were uploaded to the system to the moment the final data table was downloaded. 
The task completion times for \name{} and Baseline \texttt{A} were compared using paired Student's t-tests.

\textbf{User perceptions} were evaluated through post-task questionnaires and interviews. We used the NASA Task Load Index (6 items) to assess perceived workload, and an adapted Technology Acceptance Model (5 items) to measure system compatibility and adaptability, both using 7-point scales~\cite{kang2023synergi, wu2005drives}. Custom items gauged perceived utility in areas such as paper overview, workflow simplification, data handling, and confidence. Questionnaire data was analyzed using Wilcoxon signed-rank tests, with separate analyses for each dataset.
Qualitative feedback on system designs, workflows, and potential use cases was collected through post-study interviews and summarized to provide context and depth to our quantitative findings.

\section{Results and Analyses}
\subsection{Effectiveness of Data Extraction}

\begin{figure}[ht]
\centering
  \includegraphics[width=.95\columnwidth]{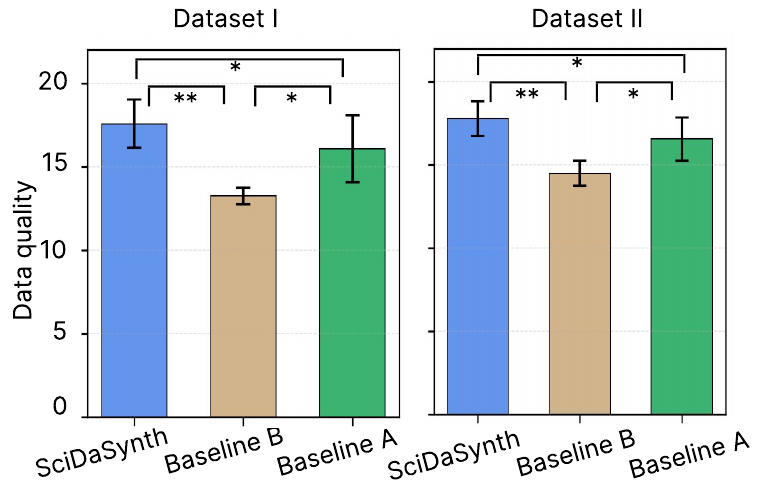}
  \caption{\rev{The data quality of using \name{}, Baseline A (human), and Baseline B (automated method). \name{} achieved the highest data quality scores for both Dataset I and Dataset II.
  *: p < 0.05, **: p <0.01.
  }}
  \label{fig:data_quality}
\end{figure}

\rev{\subsubsection{Data quality}}
As illustrated in \autoref{fig:data_quality}, \name{} consistently achieved the highest data quality scores across both datasets. For Dataset I (Nutrition Science), \name{} (M = 17.58, SD = 1.44) significantly outperformed Baseline A (M = 16.08, SD = 2.02; t = 2.34, p < 0.05) and Baseline B (M = 13.25, SD = 0.50; U = 48.0, p < 0.01). Moreover, the manual extraction method (Baseline A) also significantly outperformed the automated GPT method (Baseline B) (U = 44.0, p < 0.05), highlighting the challenges faced by fully automated systems in this domain. 
In Dataset II (Large Language Models), \name{} (M = 17.79, SD = 1.05) also significantly outperformed Baseline A (M = 16.56, SD = 1.31; t = 2.73, p < 0.05) and Baseline B (M = 14.50, SD = 0.76; U = 48.0, p < 0.01). 
Similar to Dataset I, the manual extraction method (Baseline A) significantly outperformed the automated GPT method (Baseline B) (U = 43.0, p < 0.05). 
\rev{In addition, \name{} outperformed both baselines in all data dimensions (shown in \autoref{tab:datasetI_itemwise_v1} and \autoref{tab:datasetII_itemwise}), while all three systems struggled with some error-prone fields such as raw nutrient values and units.}
These results demonstrate \name{}'s effectiveness in producing high-quality data extractions across diverse scientific domains.

\begin{figure}[ht]
\centering
  \includegraphics[width=.9\columnwidth]{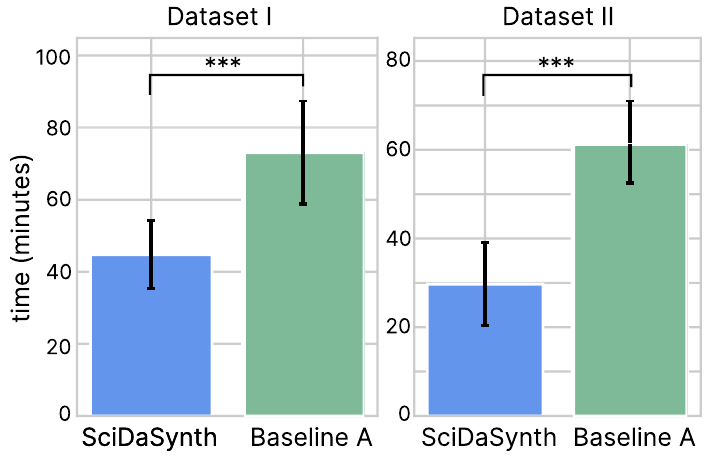}
  \caption{
  The task completion time of using \name{} and Baseline \texttt{A}.
  The pairwise comparison was significant.
  ***: p < 0.001.}
  \label{fig:time_stats}
\end{figure}

\subsubsection{Efficiency}
\autoref{fig:time_stats} illustrates the task completion time for \name{} compared to Baseline A across both datasets. For Dataset I, participants using \name{} completed the task in significantly less time (M = 44.74, SD = 9.37 minutes) compared to Baseline A (M = 73.22, SD = 14.17 minutes; t = 5.19, p < 0.001). This represents a substantial 38.9\% reduction in task completion time.
The efficiency gain was even more pronounced for Dataset II, with \name{} (M = 29.72, SD = 9.37 minutes) significantly outperforming Baseline A (M = 61.59, SD = 9.43 minutes; t = 7.38, p < 0.0001), amounting to a 51.7\% reduction in task completion time.

These findings, combining superior data quality and significantly faster task completion times, demonstrate \name{}'s capability to facilitate more efficient and accurate data extraction across various scientific domains.

\subsubsection{Case analyses of automated LLM baseline}
The automated Baseline B achieved an overall accuracy of 67.5\% (13.50/20) for Dataset I (Nutrition Science) and 73.8\% (14.75/20) for Dataset II (LLM Survey). We investigated the failure cases across both datasets and identified three major reasons for these shortcomings:

\textbf{First, incomprehensive understanding of the query in the specific paper context}. 
This issue was the most prevalent and more severe in Dataset I (Nutrition Science) due to its domain-specific terminology.
When asking about raw nutrient values in crops, Baseline \texttt{B} failed to contextualize the meaning of ``raw'' in individual paper contexts. 
For example, 
some papers might use words like ``unprocessed'' and ``unwashed'' or imply it in the tables with the processing start time equal to zero, which the system failed to recognize.
Also, there were cases where one paper could have multiple crop types, but Baseline \texttt{B} extracted only one.
In Dataset II, ``pretrained data scale'' was sometimes misinterpreted as fine-tuning dataset sizes by LLMs.

\textbf{Second, insufficient and incorrect table and figure information extraction}.
Both datasets presented challenges in this area.
In Dataset I, many failure cases stemmed from the retrieved tables and figures.
Some tables, which had very complex designs and structures (\eg, hierarchical data dimensions), were parsed wrongly. 
And some information in the figures was overlooked.
The quality of the retrieved information impacted the LLMs' reasoning, resulting in outputting ``empty'' cell values for specific dimensions.

\textbf{Third, missing associations between different parts of papers.} 
In some instances, data in tables were incomplete and required interpretation with information from other sections. For example, when asking for what crops are in a paper, the system retrieved and reported all crop variety numbers from one table instead of crop names. 
However, the corresponding crop names were recorded in method sections, demonstrating the mappings between crop names and their variety numbers.
Similarly, in Dataset II, when extracting ``model name'' and ``model size'', LLMs sometimes reported only one model type (\eg, BERT) and their sizes without identifying and differentiating the model variants (\eg, T5-base and T5-large) mentioned in different pages or sections.

\subsection{User perceptions towards \name{}}

\subsubsection{Streamline the data extraction workflow}
Participants consistently reported that \name{} significantly simplified and accelerated their data extraction process, addressing the key challenge of efficiently processing multi-modal information within scientific literature. 
The system's ability to automatically generate structured data tables from diverse sources within papers was particularly praised for its time-saving benefits.
\imp{P12} highlighted the efficiency gain, stating, ``I was impressed by how the system could extract and combine information from method descriptions, results tables, and even figures into a coherent data table. This usually takes me hours to do manually.'' This sentiment was echoed by \imp{P8}, ``The query helps me to find all of the key information, and I only need to verify them. That improves my efficiency a lot.''
Quantitative results supported these observations, with participants rating the ``effectiveness of simplifying data extraction workflow'' significantly higher for \name{} (M=5.83, SD=0.58) compared to Baseline \texttt{A} (M=4.33, SD=1.50, p=0.012).

The system's ability to understand and accurately respond to user queries was highly rated (M=6.00, SD=0.74), as was the quality of the generated data tables (M=5.50, SD=0.80). 
This interaction was deemed as ``natural'' (\imp{P9}), ``user-friendly''(\imp{P4}), and ``comfortable'' (\imp{P12}).
\imp{P7} provided a concrete example: ``When I asked about 'nutrient retention in crops after processing', the system not only extracted the relevant data but also correctly identified and populated columns for crop types, nutrient names, retention percentages, and processing methods. This would have taken me significantly longer to compile manually.''


\subsubsection{Multi-level, multi-faceted summary and standardization}
Participants reported that \name{} significantly enhanced their ability to standardize data across the paper collection compared to Baseline \texttt{A} (M=5.67, SD=0.49 vs. M=3.75, SD=1.06, p=0.005). 
The multi-level interactive visualizations was particularly effective in helping user identify and resolve data inconsistencies.
\imp{P1} recalled, ``By selecting 'crops', 'nutrients' and 'measurement units' dimensions in the scatter plot, I immediately spotted that some papers were using '$\mu$g/g' while others used 'mg/100g' for beta-carotene content in sweet potatoes. This prompted me to unify these units.''

Building upon the insights from the semantic grouping visualization, participants found the group-based standardization feature crucial for efficiently resolving identified inconsistencies. \name{} significantly facilitated this process compared with Baseline \texttt{A} (M=5.67, SD=0.83 vs. M=4.25, SD=1.36, p=0.019).
\imp{P7} described, ``After noticing a cluster in the scatter plot representing various terms for orange-fleshed sweet potatoes, I used the group-based standardization interface. Here, I could see all variations like 'Orange Sweet Potato', 'Orange-fleshed Sweet Potato', and 'OFSP' listed together. I created a standardized term 'OFSP' and dragged all variations into this group. The system then automatically updated all relevant entries in the data table.''

Overall, participants found that the integration of visual summarization and standardization tools significantly improved the efficiency and accuracy of the data standardization process.

\subsubsection{Enhanced data validation and refinement}
Participants found \name{} highly effective for locating (M=5.50, SD=0.67), organizing (M=5.92, SD=0.79), validating (M=5.17, SD=0.83), and editing data (M=5.75, SD=0.45) (all p<0.05).
The system's ability to quickly navigate to relevant parts of papers was particularly appreciated.
\imp{P4} added, ``I could easily access pdf, which is great. The option to look at context is also helpful to verify the data. For example, I easily cross-checked the data by referring to context without skimming the whole paper.''
The highlighting of important text spans in retrieved contexts proved crucial for efficient validation.
\imp{P18} shared, ``When checking model sizes for different language models, the highlighted text made it easy verify the data accuracy."

Besides, many participants praised the batch editing feature.
\imp{P22} mentioned, ``I find several clusters pointing at the same models. ...
after locating them in the table,
it was super convenient for me to edit multiple rows of data in the table at once. ''

\subsubsection{Cognitive workload and user experience}
Participants reported that \name{} significantly reduced their workload and integrated well with their existing data extraction processes. They showed a stronger interest in using \name{} (M=5.75, SD=0.97) compared to Baseline A (M=3.92, SD=1.31) in their future work (p=0.002).
The system demonstrably lowered participants' mental workload (M=3.17, SD=1.03 vs. M=4.75, SD=0.97, p=0.015) and physical workload (M=2.92, SD=1.00 vs. M=4.25, SD=1.22, p=0.034) compared to Baseline A.
Participants found \name{} highly compatible with their existing workflows. There were significant differences between \name{} and Baseline A in terms of compatibility (p=0.027) and fit with expected ways of working (p=0.005). 

Despite the addition of new visualizations and interactions, participants generally found \name{} easy to learn. \imp{P15} remarked, "The system is intuitive to use, from the query interface to the results presentation." While \name{} received a slightly higher score on the learning difficulty scale compared to Baseline A, this difference was not significant (p=0.21).

Some participants noted a brief learning curve for certain advanced features. \imp{P10} mentioned,``Operations on cluster results, like enlarging or clearing filters, and group standardization took some getting used to.'' \imp{P18} added, ``Each component is easy to learn individually, though some parts like scatter plot didn't immediately align with my habits.''

\par{}Participants also provided some valuable suggestions for system improvement.
\imp{P8} advised supporting different languages.
\imp{P5} suggested, ``I tend to make notes and comments throughout the extraction, and it may be helpful to have a field dedicated to it.''
Other suggestions mainly involve enriching table operations (\eg, change column orders (\imp{P6}), tracking data provenance and reversing back the changes (\imp{P1}, \imp{P20}).

\subsubsection{Participants remained cautious of AI-generated results}
Participants reported comparable levels of confidence in data tables built with \name{} (M=5.75, SD=0.45) and Baseline \texttt{A} (\ie, manual extraction) (M=5.08, SD=0.90, p=0.33). However, qualitative feedback revealed a more nuanced picture of user trust and usage patterns.

Researchers' trust in AI-generated results varied based on the nature of the extraction task. They expressed higher confidence in the system's ability to extract "straightforward" (\imp{P1, P3}) and "exact" data (\imp{P8}), particularly for qualitative analyses and standard metrics. \imp{P18} noted, ``For standard hardware specifications, the system was consistently accurate. But for nuanced details about model sizes or training data scales, I felt compelled to verify manually.'' 
Participants developed strategies to leverage the system's efficiency while ensuring accuracy. 
They appreciated data visualizations to get some sense of what data might look like and context verification feature. 
Some researchers, like \imp{P11}, described an approach of gradual trust-building: ``If I'm not familiar with the area, I will read by myself first. After I get familiar with the paper type and research paradigms, I will use this system more confidently.''
Interestingly, participants' trust in the system evolved through use. \imp{P20} observed, ``Initially, I was skeptical of every extracted data point. But after verifying the system's accuracy on papers I knew well, I became more confident in its capabilities for similar extraction tasks.'' This evolution was also noted among nutrition researchers. \imp{P5} said, ``As I used the system more, I learned which types of information it extracted reliably and which required more careful verification.''

\section{Discussion}\label{sec.discussion}

\subsection{Design Implications}

\subsubsection{Structured data organization beyond table}
\rev{In this work, we developed a technical framework that enables the generation of structured data tables from a  candidate pool of scientific papers in response to users’ data extraction queries.}
The structured data table helped externalize and standardize the large scale of unstructured knowledge embedded in the paper collections. 
According to the user study,
the structured data table provided a good basis for a global understanding of paper collections
and interactive visualizations of data improved awareness of data variations in different dimensions.
In the future, systems can consider other data representations beyond table format for structuring and presenting knowledge. 
For example, the mind map is a useful diagram that can visually summarize the hierarchy within data, showing relationships between pieces of the whole.
It can help users build a conceptual framework and taxonomy for paper collections, identify future research directions, and present research findings by branching out to sub-findings, implications, and recommendations.
In addition, knowledge graphs could be useful for presenting and explaining the integration of data from multiple sources.
They can also enrich data with semantic information by linking entities to concepts in an ontology, adding layers of meaning and context, and revealing hidden connections between entities.

\subsubsection{Reduce context switch by in-situ information highlighting}
To assist users in locating, validating, and refining data, 
\name{} establishes, highlights, and maintains the connections between data and relevant information in the literature. 
In the user study, participants favored the keyword highlighting in the pop-ups of relevant data contexts for corresponding rows. And they could easily access the original source PDFs for each data record.
Both of these designs helped them validate the data quality. 
However, some participants pointed out that they needed to switch different tabs to validate data tables with the source PDF content. They also desired the text highlighting in the original paper PDFs.
All of these benefits and challenges in data validation emphasize the importance of designs for reducing context switches and in-situ highlighting of information in knowledge extraction tasks.

\subsubsection{Provide analytical guidance during information extraction}
During the system exploration in the user study, some participants mentioned that they were hesitant about what questions to ask and how they should be formatted when facing paper collections that they might not be very familiar with.
The future system should provide adaptive support and guidance for users to navigate the complex information space by suggesting information questions or user interactions for initial start, follow-ups, and clarifications~\cite{PaperPlain,wang2022interactive}.
Those user questions and interaction suggestions could also be learned from users' feedback and dynamic interactions as the question-answering process progresses.






\subsubsection{Promote collaborative effort for knowledge extraction}
In this work, we designed and built an interactive system, \name{}, that facilitates users in extracting structured data from scientific literature based on LLM-generated results.
The user study showed that \name{} improved the efficiency of data extraction while presenting a comparable accuracy to the human baseline.
However, the accuracies of both systems used by individual researchers were below 90\% despite the simple questions.
There was still significant room for improvement regarding the quality of the data extracted by individuals.
This showed that data extraction from literature is a demanding and challenging task.
The system designs and workflow can further consider how to promote collaborative effort among individuals to extract and synthesize higher quality and more reliable data.

\subsection{Limitations and Future Work}
We discuss the limitations and future work based on our design and evaluation of \name{}.

The technical limitations for future work include: 
\begin{itemize}
\setlength{\itemsep}{0pt}
        \setlength{\parskip}{0pt}
    \item \textit{Improving domain context understanding.}
    Currently, we use vanilla GPT3.5/4 to build a technical pipeline for data extraction from domain-specific literature. 
    As reflected in the user study, the LLMs may still lack a deep understanding of the specialized domains and may impact users' usage and trust of the results. Therefore, future work can consider enhancing the domain knowledge and reasoning of LLMs via various approaches, such as model finetuning on domain-related articles and iterative human-in-the-loop feedback.
    \item \textit{Incorporate more methods to measure the quality of auto-generated results.} 
    We only considered the data relevance and missingness metrics to guide users' attention for cross-checking potentially low-quality data. 
    \rev{Moreover, we rely on exact string matches to highlight answer-context relationships for easier eyeballing.}
    However, errors could occur that are not captured by our metrics and may negatively impact the final data quality.
    In the future, we can develop and integrate more quantitative metrics \rev{and fuzzy matching methods} to provide users with a more comprehensive understanding of LLM performance.
\end{itemize}

The user study evaluation has pointed out future work directions:
\begin{itemize}
\item \textit{Enhance evaluation with diverse and larger user groups}. 
In this study, we evaluated our system with 24 researchers who came from nutritional science and NLP related backgrounds. 
Inviting more researchers from different disciplines would further enhance the evaluation of \name{}.

\item \textit{Conduct longitudinal study in real research scenarios}. 
The user study was conducted based on a set of predefined data extraction tasks and paper collections. However, in real research settings, participants may have interests in different data dimensions and paper topics.  
A longitudinal study of how researchers would use \name{} can further help validate and comprehensively identify the benefits and limitations of \name{}.

\rev{\item \textit{Evaluate different prompting strategies and openweight LLMs}. In this paper, we utilize GPT-4 series models to process documents and user queries. We can further extend the system by integrating openweight models, such as Llama and QWen. Moreover, evaluating prompt variations and their impacts on system performance would help us develop and optimize prompting strategies for better LLM performance.
}
\end{itemize}

\section{Conclusion}
In this paper, we developed \name{}, an interactive system to help researchers extract and structure data from scientificliterature, previously screened and eligible for a given research question, in an efficient and systematic way. Particularly, we built an LLM-based retrieval-augmented generation framework to automatically build structured data tables according to users' data questions.
Then, the system provided a suite of visualizations and interactions to guide data validation and refinement, featuring multi-level and multi-faceted data summarization and standardization support.
Through a within-subjects study with 24 researchers from nutrition and NLP domains, we demonstrated the effectiveness of data extraction via quantitative metrics and qualitative feedback.
We further discussed design implications and limitations based on the system designs and evaluation.


\acknowledgments{%
\rev{S.L.H. was supported by the NIH under award 5T32HD087137. The content is solely the responsibility of the authors and does not necessarily represent the official views of the Eunice Kennedy Shriver National Institute of Child Health and Human Development (NICHD) or the National Institutes of Health.}
}

\section*{Author Contribution}
\rev{
X.W. led ideation, technical implementation, and manuscript writing. He also co-designed the user study and contributed to data curation and results analysis.
S.L.H. led data curation, co-designed the user study, and contributed to ideation and manuscript writing.
R.S. contributed to the technical implementation, user study, and results analysis.
S.M. and F.W. provided feedback on ideation, system designs, and the manuscript.
}

\newpage
\bibliographystyle{abbrv-doi-hyperref}

\bibliography{main}

\appendix 

\definecolor{titleblockcolor}{HTML}{006494}
\definecolor{textblockcolor}{HTML}{FFFFFF}
\newenvironment{block}[2][]{
  \begin{tcolorbox}[adjusted title=#2, fonttitle={\small\bfseries}, colback={textblockcolor}, colframe={titleblockcolor}, coltitle={white}, arc=0pt,
  outer arc=0pt, left=1pt, right=1pt, fontupper=\small, #1, breakable]
}{\end{tcolorbox}}
\newtcbox{\highlight}[1][red]
  {on line, arc = 0pt, outer arc = 0pt,
    colback = #1!10!white, colframe = #1!50!black,
    boxsep = 0pt, left = 1pt, right = 1pt, top = 2pt, bottom = 2pt,
    boxrule = 0pt, bottomrule = 1pt, toprule = 1pt}

\section{Prompts}
\label{sec:appendix}
We have leveraged LLMs to perform a variety of tasks on the content of a PDF document, including extraction, structuring, description, and question-answering. 
We present the prompts as follows.

\subsection{Data Structuring}

\begin{block}{Data Structure Designs}
Given the following question, design a structured data format to represent the answer:
Question: \highlight[blue]{$\{question\}$}.\\
Your task:\\
1. Carefully analyze the question to identify ONLY the specific information explicitly requested. \\
2. Design a table structure with columns that directly correspond to the requested information.\\
3. Provide the structure in a "record" format: $[\{\{"column\_1": "value\_description", "column\_2": "value\_description"\}\}]$ \\
4. Ensure all dictionary objects in the list share the same set of columns.
5. Use clear and descriptive names for the columns.\\
6. Avoid nested structures or hierarchical data - keep everything flat.
\\

Guidelines:\\
- Choose column names that are self-explanatory and follow a consistent naming convention. \\
- Create columns ONLY for information directly mentioned or clearly implied in the question. \\
- Do NOT add columns for information that might be related but is not specifically asked for. \\
- Describe the expected values for each column (e.g., data type, format, range, units).
- For columns with multiple possible values: \\
    * Create boolean columns for each option (e.g., "has\_feature\_X", "has\_feature\_Y").
    * Or use a numeric scale to indicate presence/absence or degree (e.g., 0-5 scale).
- For columns with a limited set of possible values, list all possible options. \\
- For numerical values (e.g., size, length, weight), specify the unit of measurement if relevant. \\
- For date/time values, specify the expected format. \\
Formulate your response as a JSON object containing the designed structure.
\\

Ensure your structure capture all relevant information from the question, while also being flexible enough to accommodate various possible answers.

\end{block}

\subsection{Document Processing}

\subsubsection{Meta Information Extraction}
This prompt is used to extract the meta information in a PDF document, such as its title, abstract, year, and authors.
\begin{block}{Meta Information Extraction}
You should extract the meta information of the given paper.
This is the paper content: \highlight[blue]{$\{paper\}$}.\\

Besides, the information you need to extract includes the following keys: ``Title'', ``Abstract'', ``Year'', ``Author'', 
``Journal/Conference'', ``ISSN'', ``Volume'', ``Issue'', ``Page'', ``DOI'', ``Link'', ``Publisher'', ``Language''.
For the page, please use the format like ``12-15'', ``134-145''. If there is only one page, the format can be ``145'', ``1345''.
When there is no such information about a key, you just return the "none" as the value of the key, but you should make sure there is no such information. You should try your best to retrieve the information and reduce the occurrence of ``none''.\\

\highlight[blue]{$\{format\_instructions\}$}
\end{block}

\subsubsection{Table Identification}
This prompt is leveraged to identify and extract tables embedded within a PDF document. However, the output of this prompt is a string representation of the table data, rather than a structured tabular format such as a CSV or JSON.
\begin{block}{Table Extraction}
I will give you a page of a pdf file. 
You need first to judge whether there is any table in the page content.
Then you need to extract the original information of the table from the page content.
The following is the page content: \highlight[blue]{$\{page\_content\}$}\\

If yes, just tell me the answer through the JSON format which includes the following keys: $table\_name$ and $table\_content$.\\

Store all the JSON in a list through ``[ ]''. Besides, $table\_name$ is the Table order, such as Table 1, Table 2, and Table 3.\\

Note that you should tell me the related region of this table (raw data) from the page content without any processing in the $table\_content$.
Besides, you shouldn't output any other things (such as 'yes' or many explanations). That means, you just need to tell me the final output in JSON format in your response.

If not, just tell me ``no''.
\end{block}

\subsubsection{Table Extraction}
This prompt is leveraged to organize the tables extracted from a PDF document. Specifically, it takes the string representation of tables extracted in the previous extraction step and transforms it into a CSV format. 
\begin{block}{Table Structuring}
I will give you a table content. You need to organize it in a CSV format. This is the step:\\
(1) You should determine the column names.\\
(2) You should fill in all the data in the corresponding column and row.\\

There are some points you should pay attention to: \\
(1) Don't leave out any of the information I gave you, you should organize all my information into a table for me.\\
(2) Be careful to ``\textbackslash n''. If \textbackslash n exists, there are two kinds of scenarios. First of all, it may be too long resulting in a branch, this time the front and back are actually one and the same. If you find that \textbackslash n before and after can not form a whole, that is a nested table. the front column name is the parent column name of the back column name. At this time, you should add a parent column name. You should pay special attention when composing column names. You can use line breaks to notice which names are in a column. Here are a few different examples:\\
\textit{\scriptsize (a) \textbf{example1}: For the column name message "Tempo de estocagem (dias)\textbackslash n 0 55 90 145 180 235 280 360", you should pay special attention to the fact that there is an \textbackslash n after Tempo de estocagem (dias), so this could mean that The column names 0 55 90 145 180 235 280 360 are sub-columns of Tempo de estocagem (dias). At this point you need to organize into:
Tempo de estocagem (dias), Tempo de estocagem (dias), Tempo de estocagem (dias), Tempo de estocagem (dias), Tempo de estocagem (dias), Tempo de estocagem (dias), Tempo de estocagem (dias), Tempo de estocagem (dias), Tempo de estocagem (dias), Tempo de estocagem (dias), Tempo de estocagem (dias)
0, 55, 90, 145, 180, 235, 280, 360.
There are the column names at the previous level and column names at the next level, respectively. \\
\textbf{There are more examples of this}:
input: All-trans-b-caroteneb(mg/g DM) 13-cis-b-carotene Retention of\textbackslash nall-trans -b-carotene (\%)d\textbackslash n(mg/g DM)c(\% of total b-carotene):
thoughts: Retention of \textbackslash nall-trans -b-carotene (\%) can be thought of as turning a row instead of two columns. \textbackslash n(mg/g DM)c (\% of total b-carotene) is a sub-column, and since it can be seen that 13-cis-b-carotene has no units, (mg/g DM)c and (\% of total b-carotene) should be sub-columns of 13-cis-b-carotene. So the final column name should be organized as:
output: All-trans-b-caroteneb (mg/g DM), 13-cis-b-carotene (mg/g DM)c, 13-cis-b-carotene (\% of total b-carotene), Retention of all-trans-b-carotene (\%)d\\
(b) \textbf{example2}: Sometimes the row breaks don't necessarily represent a relationship between the column name and the subcolumn name, such as the following: TPO2 a 23 °C, 1 atm(1) \textbackslash n (mL (CNTP).m-2.dia-1). It may just be that the data is too long to be a unit. This time TPO2 a 23 °C, 1 atm(1) (mL (CNTP).m-2.dia-1) is one unit.}\\
(3) Note some of the special symbols such as ±.\\
(4) You need to ignore some special symbols, such as Unicode code point representations(e.g., /uni0394, /uni00A0).\\
(5)You should use ``'' to wrap every cell.\\
(6) Sometimes there will be redundant spaces, and you need to deal with those depending on the context. For example, there may be many spaces in ``16  ± 0.6'' due to noise, but they actually represent ``16±0.6''.\\

This is the content of my table: \highlight[blue]{$\{table\_information\}$}

Tell me the answer in JSON format, including keys ``table\_caption'' and ``table\_content'', while ``table\_content'' should be in string of CSV format.
\end{block}

\subsection{Figure Description}
This prompt is used to generate insights and descriptions for the figures that have been extracted from a PDF document. 
\begin{block}{Figure Description}
{
``\textbf{role}'': ``user'',\\
``\textbf{content}'': [\\
\llap{\hspace{1cm}} \{\\
\llap{\hspace{1cm}} ``\textbf{type}'': ``text'',\\
\llap{\hspace{1cm}} ``\textbf{text}'': f``I will give you a figure in the paper. Besides, I will also give you the caption of this figure. You should describe the data insight in this figure based on the caption. The more detailed the description, the better. This is the caption: \highlight[blue]{$\{caption\}$}.''\\
\llap{\hspace{1cm}} \},\\
\llap{\hspace{1cm}} \{\\
\llap{\hspace{1cm}} ``\textbf{type}'': ``image\_url'',\\
\llap{\hspace{1cm}} ``\textbf{image\_url}'': \\
\llap{\hspace{1cm}} \thinspace \thinspace \thinspace \{``\textbf{url}'': f``data:image/jpeg;base64,\highlight[blue]{$\{base64\_image\}$}''\}\\
\llap{\hspace{1cm}} \}\\
    ]\\
}
\end{block}

\section{User study ratings}
\begin{figure*}[!thb]
  \includegraphics[width=\textwidth]{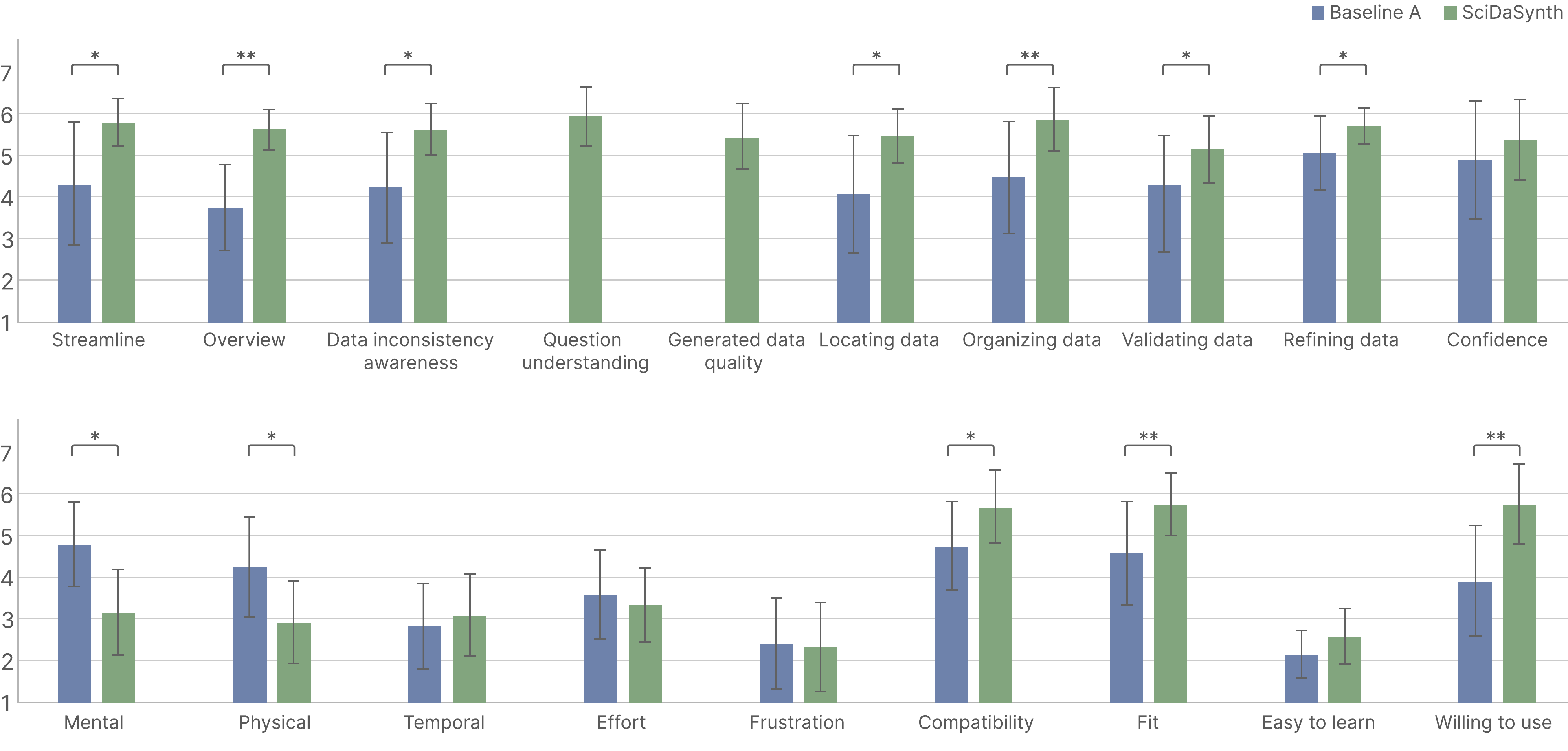}
  \caption{User study questionnaire results for both Baseline \texttt{A} and \name{}.
  The first row of items compared the ratings regarding the effectiveness in streamlining data extraction workflow, gaining an overall understanding of the paper collection, awareness of data inconsistencies, question understanding, perceived generated data quality, data locating, organization, validation, refinement, and confidence in the final data table.
  The second row compared the questionnaire items adapted from the NASA Task Load Index and the technology acceptance model.
  All ratings were on a 7-point scale.
  For ratings: ``Mental'', ``Physical'', ``Temporal'', ``Frustration'', ``Easy to learn'', the lower the ratings, the better.
  For all other ratings, the higher the ratings, the better.
  **: p < 0.01, *: p < 0.05.
  }
  \label{fig:user_ratings}
\end{figure*}

User study questionnaire ratings are presented in \autoref{fig:user_ratings}.

\section{Item-wise data extraction performance}
We also computed the performance of different systems for Datasets I and II for each dimension to be extracted. 
The dataset I results are reported in \autoref{tab:datasetI_itemwise_v1}. The dataset II results are reported in \autoref{tab:datasetII_itemwise}.
The inter-rater agreement between two raters for individual data dimensions in the two datasets by method is summarized in \autoref{tab:kappa_summary}.
\begin{table}[ht]
\centering
\caption{Dataset I: Per‐item average performance distribution across different systems.}
\label{tab:datasetI_itemwise_v1}
\resizebox{\columnwidth}{!}{%
\begin{tabular}{llccc}
\hline
\textbf{Category} & \textbf{Method} & \textbf{Correct (\%)} & \textbf{Partial (\%)} & \textbf{Incorrect (\%)} \\
\hline
\multirow{3}{*}{Crop type} 
    & SciDaSynth & 95.8 & 2.5 & 1.7 \\
    & Baseline A & 93.8 & 4.2 & 2.1 \\
    & Baseline B & 83.3 & 8.3 & 8.3 \\
\hline
\multirow{3}{*}{Micronutrient} 
    & SciDaSynth & 93.8 & 4.2 & 2.1 \\
    & Baseline A & 87.5 & 8.3 & 4.2 \\
    & Baseline B & 85.4 & 4.2 & 10.4 \\
\hline
\multirow{3}{*}{Raw value} 
    & SciDaSynth & 83.3 & 8.3 & 8.3 \\
    & Baseline A & 83.3 & 8.3 & 8.3 \\
    & Baseline B & 72.8 & 15.0 & 12.2  \\
\hline
\multirow{3}{*}{Unit} 
    & SciDaSynth & 85.4 & 8.3 & 6.3 \\
    & Baseline A & 85.4 & 7.9 & 6.7 \\
    & Baseline B & 75.0 & 12.5 & 12.5 \\
\hline
\end{tabular}
}
\end{table}

\begin{table}[ht]
\centering
\caption{Dataset II: Per‐item average performance distribution across different systems.}
\label{tab:datasetII_itemwise}
\resizebox{\columnwidth}{!}{%
\begin{tabular}{llccc}
\hline
\textbf{Category} & \textbf{Method} & \textbf{Correct (\%)} & \textbf{Partial (\%)} & \textbf{Incorrect (\%)} \\
\hline
\multirow{3}{*}{Model name} 
    & SciDaSynth & 98.8 & 0.6 & 0.6 \\
    & Baseline A & 95.0 & 3.2 & 1.8 \\
    & Baseline B & 90.6 & 6.3 & 3.1 \\
\hline
\multirow{3}{*}{Model size} 
    & SciDaSynth & 94.4 & 2.5 & 3.1 \\
    & Baseline A & 93.0 & 3.9 & 3.1 \\
    & Baseline B & 87.5 & 5.6 & 6.9 \\
\hline
\multirow{3}{*}{Pretrained data scale}
    & SciDaSynth & 90.6 & 4.4 & 5.0 \\
    & Baseline A & 89.4 & 5.6 & 5.0 \\
    & Baseline B & 81.3 & 8.8 & 10.0 \\
\hline
\multirow{3}{*}{Hardware specs} 
    & SciDaSynth & 90.0 & 8.1 & 1.9 \\
    & Baseline A & 87.5 & 5.0 & 7.5 \\
    & Baseline B & 84.4 & 5.0 & 10.6 \\
\hline
\end{tabular}
}
\end{table}

\begin{table}[ht]
\centering
\caption{Inter‐rater Cohen’s $\kappa$ by dataset and method}
\label{tab:kappa_summary}
\begin{tabular}{llc}
\hline
\textbf{Dataset}       & \textbf{Method}   & ${\kappa}$ \\
\hline
I (Nutrition)          & SciDaSynth        & 0.763 \\
                       & Baseline A        & 0.721 \\
                       & Baseline B        & 0.775 \\
\hline
II (LLM)               & SciDaSynth        & 0.738 \\
                       & Baseline A        & 0.767 \\
                       & Baseline B        & 0.842 \\
\hline
\end{tabular}
\end{table}

\end{document}